\documentclass[final,5p,times,twocolumn,letterpaper]{elsarticle}

\usepackage{natbib}
\usepackage{cancel}
\usepackage{pifont} 
\usepackage{amsthm}         
\usepackage{amsmath} 	   
\usepackage{amssymb}       
\usepackage{amsfonts}
\usepackage{graphicx} 	    
\usepackage{verbatim}
\usepackage{epsfig,bbm}
\usepackage{color}
\usepackage{colortbl}
\usepackage{float}
\usepackage{multirow}
\usepackage{lipsum}

\usepackage{subcaption}

\definecolor{babyblue}{rgb}{0.54, 0.81, 0.94}
\definecolor{corn}{rgb}{0.98, 0.93, 0.36}

\begin{document}

\begin{frontmatter}

\title{Nearly scale-invariant curvature modes
from entropy perturbations during graceful exit}

\author[add1,add2]{Anna Ijjas\corref{cor1}}
\ead{anna.ijjas@aei.mpg.de}
\cortext[cor1]{Corresponding author}
\author[add0]{Roman Kolevatov}

\address[add1]{Max Planck Institute for Gravitational Physics (Albert Einstein Institute), 30167 Hannover, Germany}
\address[add2]{Institute for Gravitational Physics, Leibniz University Hannover, 30167 Hannover, Germany}
\address[add0]{Department of Physics, Princeton University, Princeton, NJ 08544, USA}

\date{\today}

\begin{abstract}
In this Letter, we describe how a spectrum of entropic perturbations generated during a period of 
slow contraction can source a nearly scale-invariant spectrum of curvature perturbations on length scales larger than the Hubble radius during the transition from slow contraction to a classical non-singular bounce (the `graceful exit' phase).  The 
sourcing occurs naturally through higher-order scalar field kinetic terms common to classical (non-singular) bounce mechanisms.   We present a concrete example in which, by the end of the graceful exit phase, 
 the initial entropic fluctuations have become negligible and the curvature fluctuations have a nearly scale-invariant spectrum with an amplitude consistent with observations. 

 \end{abstract}


\end{frontmatter}

\section{Introduction.} 

Observational evidence \cite{Komatsu:2008hk,Sievers:2013ica,Aghanim:2018eyx} combined with theoretical reasoning \cite{Bardeen:1983qw,Bedroya:2019snp} strongly indicate that the gravitationally bound structures (galaxies, galaxy clusters, {\it etc.}) that comprise our universe originate from quantum fluctuations of scalar fields generated on sub-Hubble wavelengths that evolve to induce classical curvature perturbations with a nearly scale-invariant and gaussian spectrum on super-Hubble wavelengths.  According to the leading paradigms, the relevant quantum fluctuations are generated
during a primordial smoothing phase 
at energy densities sufficiently 
below the Planck density so that the cosmological background can be described to leading order by classical 
equations of motion. 

Two candidates for the smoothing phase are 
a period of accelerated expansion ($\dot{a}, \ddot{a}>0$) and a period of slow contraction ($\dot{a}, \ddot{a}<0$), where the spacetime geometry during the smoothing phase is well-described by the  Friedmann-Robertson-Walker (FRW) metric with scale factor $a(t)$ and the  dot denotes differentiation with respect to the physical FRW time coordinate $t$. The key underlying idea is that, in either case,   the scale factor $a(t)$ and the Hubble radius $|H^{-1}|  \equiv |a/\dot{a}| $ evolve at different rates,
\begin{equation}
\label{smoothing}
|H^{-1}| \propto a^{\epsilon},
\end{equation}
 as determined by the equation of state 
 \begin{equation}
 \label{eos}
 \epsilon \equiv \frac32 \left(1+ \frac{p}{\varrho}\right)
 \end{equation}
of the dominant stress-energy component with pressure $p$ and energy density $\rho$ \cite{Ijjas:2018qbo}. 
During an accelerated expansion ($\epsilon<1$) phase, the Hubble radius stays nearly constant while scalar field fluctuation wavelengths, which grow in proportion to the scale factor $a$, stretch at an ultra-rapid rate to become super-Hubble.  In contrast, the Hubble radius during slow contraction ($\epsilon>3$) shrinks ultra-rapidly while the scale factor is nearly constant.  For example,
in a typical slow contraction phase, the Hubble radius might shrink by a factor of $2^{50}$ during which the scale factor decreases by only a factor of two \cite{Ijjas:2019pyf}.   As a result, fluctuation wavelengths that were sub-Hubble at the beginning of the phase evolve to become super-Hubble by the end. 

However, generating scalar field fluctuation modes with super-Hubble wavelengths is necessary but not sufficient to explain cosmological observations. To explain measurements of the cosmic microwave background and the power spectrum of gravitationally bound structures, the scalar field fluctuations must somehow source a nearly scale-invariant spectrum of  co-moving curvature fluctuations of the metric with the amplitude of $\sim 10^{-5}$. 

In general, scalar field fluctuations source two types of metric fluctuations: {\it adiabatic} fluctuations on constant mean curvature hypersurfaces; and {\it entropic} fluctuations on hypersurfaces of constant energy density \cite{Bardeen:1980kt,Mukhanov:1990me}. 
Notably, scalar fields in backgrounds undergoing accelerated expansion can generate both types.  If it can be arranged that the metric fluctuations are purely adiabatic and of the correct small amplitude, they can potentially account for the observed temperature fluctuations of the cosmic microwave background. However, accelerated expansion also inevitably stretches rare, large-amplitude scalar field fluctuations that source large-amplitude adiabatic fluctuations of the metric.  These 
large metric fluctuations trigger the well-known quantum runaway problem, an effect that spoils the spectrum and destroys homogeneity and isotropy altogether \cite{Steinhardt:1982kg,Vilenkin:1983xq,Guth:2000ka}.
Slow contraction, on the other hand, can only amplify entropic modes.  Adiabatic modes (as well as gravitational waves \cite{Boyle:2003km}) experience a growing anti-friction due to the rapidly decreasing Hubble radius which leads to their decay.  This eliminates the quantum runaway problem, an important and distinctive advantage of the slow contraction scenario. To date, we do not know of any other smoothing mechanism that could do the same.

It is well-known that, during smoothing slow contraction, a non-linear sigma type kinetic interaction between two scalar fields naturally leads to a nearly scale-invariant and gaussian spectrum of super-Hubble relative field fluctuations -- purely entropic modes --
which are quantum generated long before the modes leave the Hubble radius \cite{Li:2013hga,Fertig:2013kwa,Ijjas:2014fja,Levy:2015awa}. 
In this Letter, we demonstrate how these entropic modes can source 
curvature perturbations on super-Hubble scales during the  `graceful exit' phase,  {\it i.e.}, the transition from slow contraction to the bounce stage.  
We show that the sourcing is due to a common feature of classical (non-singular) bounce models in which  higher-order kinetic terms associated with the scalar matter fields become important during graceful exit \cite{Ijjas:2016tpn,Ijjas:2016vtq}.  For concreteness, we present an example in which the only significant fluctuations at the beginning of the graceful exit phase are entropic but, by the end of the phase, 
 the entropic fluctuations have become negligible and the curvature fluctuations have a nearly scale-invariant spectrum with an amplitude consistent with observations.

\section{Cosmological model}
\label{sec:setup}
In the scenario that we shall consider the cosmological evolution is sourced by two kinetically-coupled scalar fields $\phi$ and $\chi$ both of which are minimally coupled to Einstein gravity. 
The corresponding Lagrangian density is defined as
\begin{equation}
\label{eq:setup_action}
{\cal L}= {\textstyle \frac12} R - {\textstyle \frac12}(\partial_\mu\phi)^2 
- {\textstyle \frac12}\Sigma_1(\phi)(\partial_\mu\chi)^2
+{\textstyle \frac14}\Sigma_2(\phi)(\partial_\mu\chi)^4 - V(\phi,\chi),
\end{equation}
where $R$ is the Ricci scalar; $\Sigma_1(\phi)$ is the quadratic kinetic coupling function; $\Sigma_2(\phi)$ is the quartic kinetic coupling function and $V(\phi,\chi)$ is the scalar potential depending on both $\phi$ and $\chi$ fields. The potential is steep and negative along the  $\phi$ direction while nearly constant along the $\chi$ direction. Throughout, we use reduced Planck units.

Our interest here is analyzing what happens after a period of slow contraction has already homogenized and isotropized 
spacetime well-described by an FRW metric.  Varying the action given through Eq.~\eqref{eq:setup_action} with respect to the scalar fields and evaluating for the FRW background yields the evolution equations for $\phi$ and $\chi$:
\begin{subequations}
\label{eq:setup_eom_field}
\begin{align}
\label{eq:setup_eom_field_phi}
&\; \ddot{\phi} + 3H\dot{\phi} + V_{,\phi} = \left(\Sigma_{1,\phi} + {\textstyle \frac12}\Sigma_{2,\phi}\dot{\chi}^2\right){\textstyle \frac12}\dot{\chi}^2
,\\
\label{eq:setup_eom_field_chi}
&\left( 1 + \frac{2\Sigma_2\dot{\chi}^2}{\Sigma_1 + \Sigma_2\dot{\chi}^2}\right) \,\ddot{\chi} 
 +  \left(3H + \frac{\Sigma_{1,\phi} + \Sigma_{2,\phi}\dot{\chi}^2}{\Sigma_1 + \Sigma_2\dot{\chi}^2} \dot{\phi}\right)\,\dot{\chi} =
\\
& =  - \frac{V_{,\chi}}{\Sigma_1 + \Sigma_2\dot{\chi}^2} \,.
\nonumber
\end{align}
\end{subequations}
Note that the symmetries of the FRW space-time geometry lead  to  spatially homogeneous background field distributions, {\it i.e.}, $\phi = \phi(t)$, $\chi = \chi(t)$.

Variation of Eq.~\eqref{eq:setup_action} with respect to the metric yields the stress-energy tensor $T^\mu{}_\nu$. On an FRW background, scalar fields (collectively) act as perfect fluids and can be associated with an energy density $\rho$ and pressure $p$ being given by the temporal and spatial components of $T^\mu{}_\nu$:
\begin{subequations}
\label{eq:background_p_rho}
\begin{alignat}{2}
\rho  & = -T^0{}_0 &&= {\textstyle \frac12}\dot{\phi}^2 
+ {\textstyle \frac12}\left(\Sigma_1 + {\textstyle \frac32}\Sigma_2\dot{\chi}^2\right)\dot{\chi}^2 + V
,\\
p &= {\textstyle \frac13} T^i{}_i&&={\textstyle \frac12}\dot{\phi}^2 
+ {\textstyle \frac12} \left(\Sigma_1 + {\textstyle \frac12}\Sigma_2\dot{\chi}^2\right)\dot{\chi}^2 - V.
\end{alignat}
\end{subequations}
With Eq.~\eqref{eos}, we can define the effective equation of state associated with the `fluid' as follows: 
\begin{equation}
 \epsilon = 3 -   \frac14\dfrac{\Sigma_2\dot{\chi}^4}{H^2} - \dfrac{V}{H^2}.
\end{equation}

Finally, the Friedmann constraint and evolution equation take the form:
\begin{subequations}
\label{eq:setup_eom_gravity}
\begin{align}
3H^2 &= \rho = {\textstyle \frac12}\dot{\phi}^2 + {\textstyle \frac12} \left(\Sigma_1 + {\textstyle \frac32}\Sigma_2\dot{\chi}^2\right)\dot{\chi}^2 + V,\\
\label{eq:second_fried}
-2\dot{H} &= \rho + p =  \dot{\phi}^2 + \left(\Sigma_1 + \Sigma_2\dot{\chi}^2\right)\dot{\chi}^2.
\end{align}
\end{subequations}

\section{Entropy Modes from Slow Contraction}
\label{sec:slow}

As an example, we consider a scalar field potential that, during the slow contraction phase, 
 is negative and steeply graded along the $\phi$ 
direction:
\begin{equation}
\label{eq:slow_contr_potential}
V(\phi, \, \chi) \approx -V_0e^{\phi/M}.
\end{equation}
Here $V_0>0$ is constant and $M$ is the characteristic mass scale associated with $\phi$.
At low energies, especially during the smoothing slow contraction phase, higher-order kinetic terms as well as the $\chi$ field's potential energy density are negligible, such that the Einstein-scalar system reduces to the simple set of evolution and constraint equations:
\begin{subequations}
\label{eq:setup_eom_field-s}
\begin{align}
\label{eq:setup_eom_field_phi-s}
&\ddot{\phi} + 3H\dot{\phi} - {\textstyle \frac{V_0}{M}} e^{\phi/M} \approx 0
,\\
\label{eq:setup_eom_field_chi-s}
&\ddot{\chi} 
 +  \left(3H + \frac{\Sigma_{1,\phi} }{\Sigma_1 } \dot{\phi}\right)\,\dot{\chi} 
\approx  0 \,,\\
& 3H^2  \approx  {\textstyle \frac12}\dot{\phi}^2 +  {\textstyle \frac12}\Sigma_1 \dot{\chi}^2 - V_0 e^{\phi/M}.
\end{align}
\end{subequations}
For $\Sigma_1 = e^{\phi/m}$, where $m\lesssim M$, it is straightforward to show (see Ref.~\cite{Levy:2015awa}) that the unique attractor
scaling solution of the Einstein-scalar system of equations~\eqref{eq:setup_eom_field-s} is:
\begin{subequations}
\begin{align}
\label{eq:slow_contr_phi_sol}
\phi \approx -2M\times\ln\left(-At\right),\quad
&a \approx \left(-t\right)^\frac{1}{\epsilon},\quad
\epsilon \approx {\textstyle \frac12} \left(\frac{M_{\rm Pl}}{M}\right)^2,\\
&\dot{\chi} \approx 0,
\end{align}
\end{subequations}
where $A = M_{\rm Pl}^{-1} \epsilon\,\sqrt{V_0/\left(\epsilon-3\right)}$ and 
 $\epsilon$ is the equation of state as defined in Eq.~\eqref{eos}. The FRW time coordinate $t$ runs from large negative to small negative values and we normalized $a$ such that $a=1$ at the onset of slow contraction. 
 Apparently, the solution~\eqref{eq:slow_contr_phi_sol} is corresponding to   the physical situation of smoothing slow contraction with $|H^{-1}| \approx a^{\epsilon}$ shrinking exponentially faster than the scale factor $a$. For example, for $M/M_{\rm Pl}\sim0.1$, $\epsilon \sim 50$ such that $|H^{-1}|$ shrinks by a factor of $2^{50}$ while $a$ decreases by a factor of 2,  the case described in the Introduction. 
 
 A key to the background dynamics during the smoothing phase is the non-linear $\sigma$-type kinetic interaction $\Sigma_1(\phi)(\partial_{\mu}\chi)^2$ between the $\phi$ and $\chi$ fields.  As can be seen from Eq.~\eqref{eq:setup_eom_field_chi-s}, this contribution changes  the Hubble anti-friction term ($\propto 3H$) into a friction term
 \begin{equation}
 \label{damping}
3H + \frac{\dot{\phi}}{m} \approx \frac{2}{(-t)}\left(- 3 \left(\frac{M}{M_{\rm Pl}}\right)^2 + \frac{M}{m} \right) \gg 0,
\end{equation}
if $M_{\rm Pl}/M > 3 (m/M_{\rm Pl})$.
As a result, the two scalar fields exhibit very different dynamics. The $\chi$ field is being continuously damped by the friction in Eq.~(\ref{eq:setup_eom_field_chi-s}) until it eventually `freezes' at some constant value $\chi_0$.
At the same time, the $\phi$ field, which only experiences anti-friction according  Eq.~(\ref{eq:setup_eom_field_phi-s}), 
is 
being blue-shifted due to the pure Hubble anti-friction and hence keeps rolling down its negative potential energy curve, rapidly becoming the dominant stress-energy component, which robustly homogenizes and isotropizes the cosmological background \cite{Ijjas:2020dws}.

While the $\chi$-field does not contribute to the background smoothing, it plays an important role at perturbative order: quantum fluctuations in the $\phi$ field, which experience the same Hubble anti-friction as the background, blue-shift. The opposite is true for the $\chi$ field. Due to the modified damping term as given in Eq.~\eqref{damping}, quantum fluctuations in the $\chi$ field `see' a de-Sitter-like background and red-shift. Consequently, by means analogous to the case of inflation, they can lead to a nearly scale-invariant spectrum of $\chi$-fluctuations 
with  super-Hubble wavelengths.

This becomes particularly clear if we follow the evolution of the canonically normalized perturbation variable $v_\chi = a\sqrt{\Sigma_1}\delta\chi$, where $\delta \chi$ is the linearized field variable associated with $\chi$. For each Fourier mode with wavenumber $k$, the corresponding  Mukhanov-Sasaki equation takes the simple form:
\begin{equation}
\label{eq:ekp_perturb_vs}
v_\chi'' + \left(k^2-\frac{z''}{z}\right)v_\mathcal{\chi} = 0,
\end{equation}
where $z \equiv \sqrt{\Sigma_1} a$, and prime denotes differentiation w.r.t. the conformal time coordinate $\tau$ defined through $d\tau = a^{-1} dt$. 
Evaluating for the scaling solution as given in Eq.~\eqref{eq:slow_contr_phi_sol}, we find the variable $z$ as a function of $\tau$:
\begin{equation}
z \propto (-\tau)^{-\frac{1}{\epsilon -1}\Big( \frac{M}{m} \, \epsilon - 1 \Big)}\,,
\end{equation}
and, hence, $z''/z \propto 1/\tau^2$ turning Eq.~\eqref{eq:ekp_perturb_vs} into a Bessel equation.
At the onset of slow contraction, the energy density $\sim H^2$ is small and space-time is leading-order classical, such that it is natural to assume Bunch-Davies boundary conditions ($v_\chi = e^{-ik\tau}/\sqrt{2k}$ for $\tau \to - \infty$). 
The corresponding solution to the Bessel equation~\eqref{eq:ekp_perturb_vs} then takes the well-known form:
\begin{equation}
 v_\chi= \sqrt{\dfrac{\pi}{4}(-\tau)}H_\nu^{(1)}\,(-k\tau),
\end{equation}
where $H_\nu^{(1)}$ is a Hankel function of the first kind, and 
\begin{equation}
\nu^2 = \frac{1}{4} + \tau^2\frac{z''}{z} = \frac14 \left( 1+ 2\,\frac{\frac{M}{m}\epsilon - 1}{\epsilon - 1} \right)^2.
\end{equation}
On large scales ($-k\tau \ll 1$), the modes  have the following asymptotic:
\begin{equation}
\label{eq:ekp_vs}
v_\chi \propto \left(-\tau\right)^{-\nu-\frac{1}{2}}\cdot k^{-\nu}.
\end{equation}
such that the spectral tilt of the $\chi$ perturbations is given by
\begin{equation}
\label{eq:nu_ns_final}
n_s - 1 = 3 - 2\nu =2\left( 1 - \frac{\frac{M}{m}\epsilon - 1}{\epsilon - 1} \right).
\end{equation}
Note that strictly equal mass scales ($M=m$) lead to an exactly scale-invariant spectrum ($n_s - 1= 0$). However, if the mass $M$ is slightly greater than the scale $m$ associated with the kinetic interaction ({\it e.g.,} $M =1.02 m$), the spectrum is slightly red in agreement with microwave background observations ($n_s -1 \simeq -0.04$).

It has been common to identify fluctuations in $\chi$ with entropy perturbations because, in field space, the $\chi$ field defines a direction perpendicular to the adiabatic background trajectory; see, {\it e.g.}, Ref.~\cite{Gordon:2000hv}. However, this geometric interpretation has limited applicability: it is only valid in cases where all scalar matter fields  have canonical kinetic energy density; see Ref.~\cite{Ijjas:2020cyh}. A more general and precise statement is that
 $\delta \chi$ sources (macroscopic) entropy modes,
\begin{equation}
\label{def-S-hydro}
\mathcal{S} \equiv  H\left(\dfrac{\delta p}{\dot{p}} - \dfrac{\delta \rho}{\dot{\rho}}\right) \equiv H \frac{ \delta p_{\rm nad}}{\dot{p}},
\end{equation}
provided it generates a non-zero pressure contribution on hypersurfaces of constant density, {\it i.e.}, 
 \begin{equation}
 \label{dpnad}
 \delta p_{\rm nad} \equiv \delta p - \frac{\dot{p}}{\dot{\rho}}\delta\rho \neq 0.
 \end{equation}
 As we will show next, this occurs naturally during graceful exit from slow contraction to the onset of the bounce stage. Furthermore, we demonstrate that the same entropy modes source super-Hubble curvature modes consistent with cosmic microwave background observations.

\section{Sourcing Curvature Modes during Graceful Exit}
\label{sec:source}
The smoothing slow contraction phase comes to an end when the scalar field kinetic energy increases relative to the 
potential energy such  that $\epsilon \rightarrow 3$. The phase that connects to the bounce stage is called `graceful exit.' In scenarios where the cosmological bounce occurs at high yet sub-Planckian energies, this intermediate stage is dominated by the kinetic energy of the fields. In particular, this is precisely where one expects higher-order kinetic terms to start playing a role; see, {\it e.g.} Refs.~\cite{Ijjas:2016tpn,Ijjas:2016vtq,Ijjas:2017pei}. As we will see, this naturally leads to the sourcing of super-Hubble curvature modes by the fluctuations in $\chi$ generated during the smoothing phase.

In Ref.~\cite{Ijjas:2020cyh}, we have shown that, on large scales ($k\ll a|H|$), the conservation of stress-energy leads to a simple relation describing the evolution of curvature fluctuations ${\cal R}$ as a function of the entropy modes: 
\begin{equation}
\label{eq:sourcing_R}
\dot{\cal R} \approx -3H\frac{\dot{p}}{\dot{\rho}} \mathcal{S} = H \frac{\delta p_{\rm nad} }{\rho + p}.
\end{equation}
In spatially-flat gauge, the co-moving curvature perturbation can be expressed as a function of the perturbed scalars as follows:
\begin{equation}
\label{eq:sourcing_R_F}
\mathcal{R} \equiv H\dfrac{\dot{\phi}\delta\phi + \sqrt{\Sigma_1 + \Sigma_2\dot{\chi}^2}\dot{\chi}\delta\chi}{\dot{\phi}^2 + \left(\Sigma_1+\Sigma_2\dot{\chi}^2\right)\dot{\chi}^2},
\end{equation}
and, for our model as described in Eq.~\eqref{eq:setup_action}, the non-adiabatic pressure contribution is given by
\begin{align}
\label{eq:perturbs_non-adiabatic_press}
\delta p_{\rm nad} 
\approx 2c_S^2 &\Big[ - \Sigma_2\dot{\chi}^3 \dot{\phi} \times \dot{\cal F}
\\
+&
 \Big( \big(\Sigma_1 + \Sigma_2\dot{\chi}^2\big)\big(V_{,\phi} + {\textstyle \frac14}\Sigma_{2,\phi}\dot{\chi}^4\big)\,\dot{\chi}  - V_{,\chi}\dot{\phi} - \Sigma_2\dot{\chi}^3 \ddot{\phi} \Big) \times {\cal F}\Big]
\,.\nonumber
\end{align}
Here 
\begin{equation}
{\cal F} \equiv \left( \frac{\dot{\phi} \dot{\chi}}{\dot{\phi}^2 + \Sigma_1\dot{\chi}^2 + \Sigma_2\dot{\chi}^4}\right) \left( \frac{\delta\chi}{\dot{\chi}} - \frac{\delta\phi}{\dot{\phi}}\right)
\end{equation}
describes the relative field fluctuations; and the formal quantity
\begin{equation}
c_S^2 \equiv \frac{{\dot{\phi}^2 + \Sigma_1\dot{\chi}^2 + \Sigma_2\dot{\chi}^4} }{{\dot{\phi}^2 + \Sigma_1\dot{\chi}^2 + 3\Sigma_2\dot{\chi}^4} }\end{equation}
denotes the propagation speed of the adiabatic mode. The detailed dynamics of ${\cal R}$ and ${\cal S}$ can be determined by integrating the closed system of Eqs.~(\ref{eq:app_constraints}-\ref{eq:app_eom}) under spatially-flat gauge conditions, as detailed in the Appendix. 

During slow contraction, $\delta p_{\rm nad}  \approx 0$, as can be seen when evaluating Eq.~\eqref{eq:perturbs_non-adiabatic_press} for the scaling attractor solution for which  $\dot{\chi} \approx 0$. 
During graceful exit, on the other hand, $\dot{\chi}$ is non-zero and $\Sigma_2$ is non-negligible. In contrast to the smoothing phase, the relative field fluctuations ${\cal F}$ lead to a non-zero non-adiabatic pressure which, in turn, sources super-Hubble co-moving curvature modes.

\begin{figure}[tb]
\center{\includegraphics[width=1\linewidth]{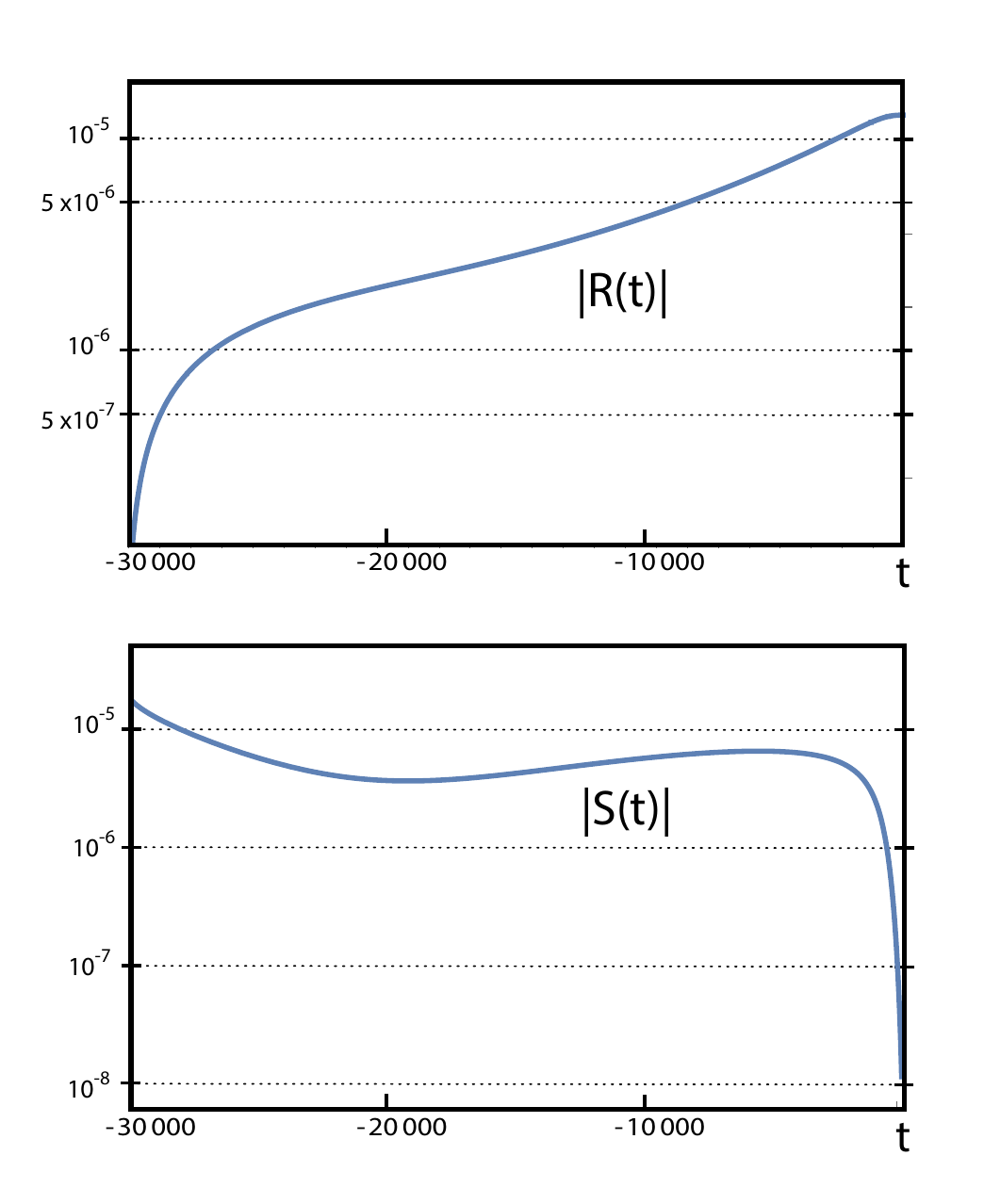}} 
\caption{A plot of the magnitude of the curvature perturbation on co-moving hypersurfaces, $|{\cal R}(t)|$ (top),  and the entropy perturbation, $|{\cal S}(t)|$ (bottom), as a function of time $t$  (expressed in 
reduced Planck units) for the example discussed in the 
text.} 
\label{Fig:R2Figure1}
\end{figure}
Since the curvature perturbations ${\cal R}$ are sourced by ${\cal S}$ on super-Hubble scales, as indicated in Eq.~\eqref{eq:sourcing_R}, ${\cal R}$ automatically inherits the nearly scale-invariant form of the entropic spectrum. 
\begin{figure}[tb]
\center{\includegraphics[width=1\linewidth]{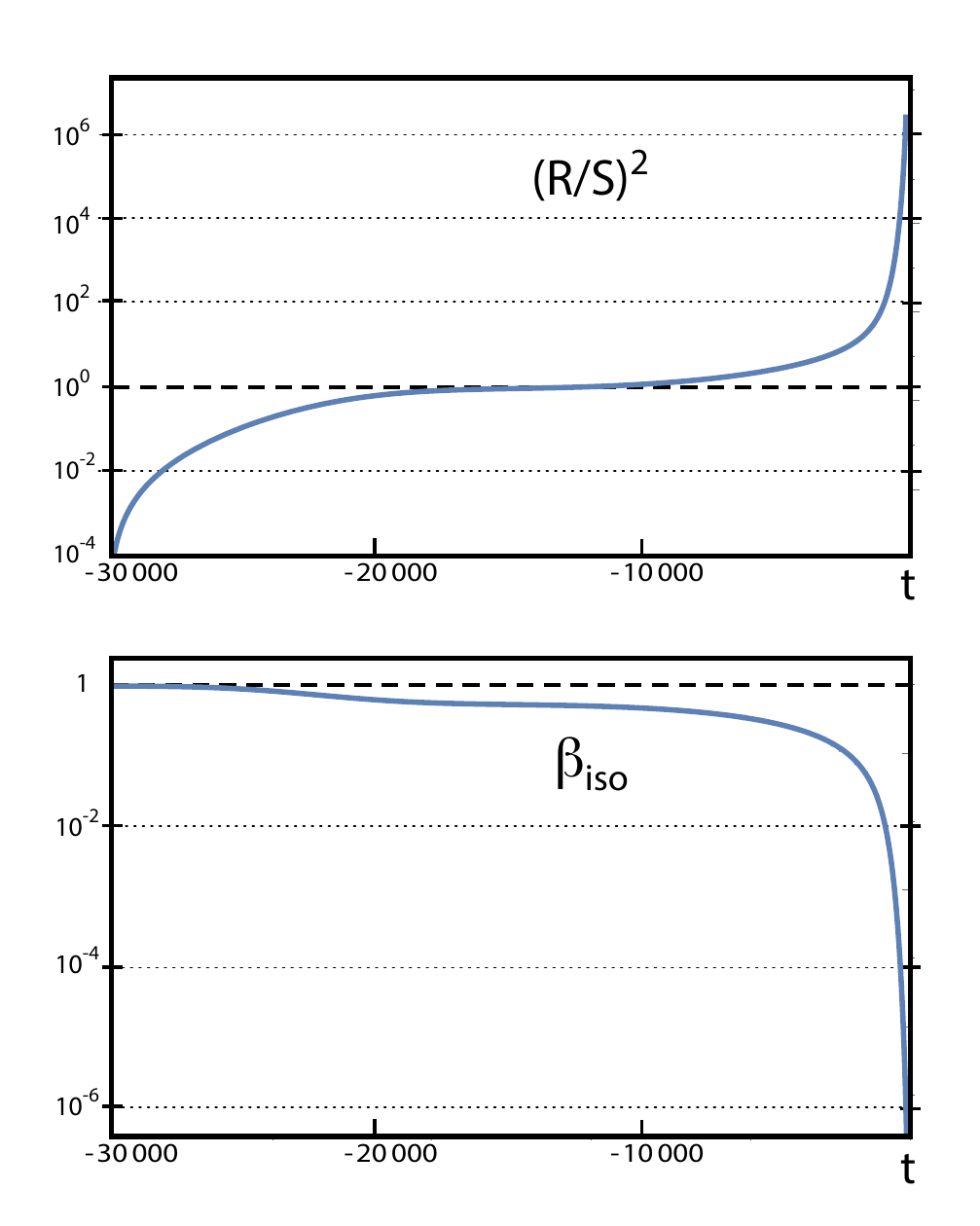}} 
\caption{ A plot of $[{\cal R}(t)/{\cal S}(t)]^2$ (top) for the example discussed in the text.   
At the end of the slow contraction phase and entering the bounce phase, $|{\cal R}| \ll {\cal S}$ is negligible; but, by the time the bounce 
would occur ($t\approx -100$ in this example), the sourcing of the curvature perturbation by the entropic perturbation
leads to $|{\cal R}| \gg {\cal S}$.  Consequently, the fractional contribution of the entropy modes to the total power spectrum,
$|\beta_{\text{iso}}(t)| \equiv {\cal S}^2/({\cal R}^2+{\cal S}^2)$, is nearly one entering the bounce phase but negligibly small by the time the bounce occurs, consistent with current observations. }
\label{Fig:R2Figure2}
\end{figure}

{
A particular example illustrating the sourcing of ${\cal R}(t)$ by ${\cal S}(t)$ is presented in Fig.~\ref{Fig:R2Figure1}. 
In this example, the kinetic coupling functions in the action, 
Eq.~(\ref{eq:setup_action}), have a simple exponential form: $\Sigma_1=e^{\phi/m}$, $\Sigma_2= e^{\phi/m_2}$, with $m =0.67$ and $m_2=-0.5$. 
We have also taken the dependence of the potential on $\phi$ in this transition phase after slow contraction to be negligible and on $\chi$ to be small:  $V(\phi, \chi) = V_0 \chi$, with $V_0 = 2 \times 10^{-13}$ such that $V(\phi, \chi)$ is small compared to the total kinetic energy density throughout the bounce phase (from the end of slow contraction at time $t=t_i = - 3 \times 10^4$ to the bounce itself at $t=t_f = -10^2$, expressed in reduced Planck time units), as expected when approaching a bounce. The background initial conditions are the following: $\phi_i=-7.5$, $\dot{\phi}_i=2.7 \cdot 10^{-5}$, $\chi_i=0$, $\dot{\chi}_i=-2.3 \cdot 10^{-6}$. These background conditions were chosen such that the energy density in $\phi$ dominates over the energy density in $\chi$  at the end of the slow contraction phase, as expected in bouncing scenarios:  that is $\Omega_\phi/\Omega_\chi |_{t=t_i} \gg 1$. Conversely, the initial ratio of the curvature perturbation is set to be negligible  compared to the entropic perturbation ($({\cal R}/{\cal S})^2 \approx   10^{-4} \ll 1$), as expected when first ending the slow contraction phase because the adiabatic fluctuations are not amplified during slow contraction.  

The evolution of ${\cal R}(t)$ and ${\cal S}(t)$ for a super-Hubble radius mode with $k/a|H| \ll1$ is shown in Fig.~\ref{Fig:R2Figure1} as obtained from numerical integration of Eqs.~(\ref{eq:app_constraints}-\ref{eq:app_eom}) in the spatially-flat gauge.  
As anticipated, beginning from a negligible curvature perturbation, the sourcing of ${\cal R}$ by the entropic perturbation causes ${\cal R} \ll 10^{-5}$ to grow to an amplitude consistent with observations (see Fig.~\ref{Fig:R2Figure2}),   ${\cal R} \approx  10^{-5}$.
The total curvature perturbation power spectrum amplitude is then
\begin{equation}
\left<\mathcal{R}^2\left(x\right)\right> = \int \dfrac{d^3 k}{\left(2\pi\right)^3}\left(\dfrac{\mathcal{R}}{k^{\nu}}\right)^2 =  \int \dfrac{d k}{k}\left(\dfrac{\mathcal{R}^2}{2\pi^2}\right) k^{3-2\nu} =  \int \dfrac{d k}{k}\Delta^2_\mathcal{R}\left(k\right).
\end{equation} 
Taking the ${\cal R}$ obtained from the numerical integration to correspond to $k_*=0.002~Mpc^{-1}$ , we obtain 
\begin{equation}
\label{eq:source_final}
\Delta^2_\mathcal{R}\left(k\right)|_{t=t_f} = 2.5 \cdot 10^{-9}\quad\text{with}\quad n_s \simeq 0.96,
\end{equation}
in accord with current observations \cite{Komatsu:2008hk}.  
Over the same period of evolution, Fig.~\ref{Fig:R2Figure2} shows that $({\cal R}/{\cal S})^2$ grows to be greater than $10^6$ such that the fractional contribution of isocurvature perturbations to the total power spectrum $\beta_{\text{iso}}$  becomes negligible, also in accord with current observations.
}

\section{Conclusion}
\label{sec:concl}
Scalar field perturbations of quantum origin can source adiabatic or entropic fluctuations on super-Hubble scales. Entropic fluctuations on super-Hubble scales can source curvature modes on super-Hubble scales as a consequence of stress-energy conservation. If our universe has undergone a phase of slow contraction that connects to the current expanding phase through a cosmological bounce, adiabatic and gravitational wave fluctuations from the smoothing phase decay and therefore cannot contribute to
the observed fluctuation spectra of the cosmic microwave background. Rather, we would expect that the temperature anisotropies stem from super-Hubble entropy modes generated during slow contraction that sourced curvature modes before the onset of decelerated expansion.

In this paper, we described a scenario for how this mechanism might naturally occur during graceful exit when slow contraction ended but the bounce has not yet occurred. Furthermore, we presented an explicit example that generates a spectrum of primordial perturbations that agrees with current cosmological observations.

The key ingredients of this new mechanism are:
\begin{enumerate}
\item[-] a non-linear $\sigma$-type kinetic interaction between two scalar fields that acts as a friction term (typical of de Sitter-like expansion) on one of the fields which it `freezes,' leading to a nearly scale-invariant spectrum of this field's quantum fluctuations;
\item[-] a higher-order quartic kinetic term that typically comes to dominate at the end of slow contraction and at the onset of the classical (non-singular) bounce stage. This term naturally leads to a non-adiabatic pressure contribution, sourcing super-Hubble curvature modes before the bounce occurs.
\end{enumerate}
This novel kinetic sourcing mechanism opens up several new avenues for future research. For example, it will be interesting to see if different graceful exit and bounce mechanisms leave different detectable imprints on the spectrum when the curvature modes are being sourced by entropy modes during graceful exit. 

\vspace{.1in}
{\it Acknowledgements}
We thank Paul J. Steinhardt for helpful comments and discussions. The work of A.I. is supported by the Lise Meitner Excellence Program of the Max Planck Society and by the Simons Foundation grant number 663083.
R.K. thanks the Max Planck Institute for Gravitational Physics (Hannover) for hospitality, where parts of this work were completed.

\appendix
\section{Evolving the linearized scalars $\delta\phi$ and $\delta\chi$}
\label{sec:appendix}

In this Appendix, we derive the evolution equations for the perturbed scalars $\delta\phi$ and $\delta\chi$ that we used to numerically compute the example presented in Figs.~\ref{Fig:R2Figure1}~and~\ref{Fig:R2Figure2} above.

Scalar variables of the linearly perturbed line element for a spatially-flat Friedmann-Robertson-Walker (FRW) space-time are given by
\begin{eqnarray}
\label{eq:app_metric}
ds^2 &=& -(1 + 2\alpha)dt^2 +2a\left(t\right)\partial_i\beta dt\mathrm{d}x^i 
\\
&+& a^2\left(t\right)\left[\left(1 - 2\psi\right)\delta_{ij} + 2\partial_i\partial_jE\right]dx^idx^j,\nonumber
\end{eqnarray}
where $\alpha$ and $\beta$ are the linearized lapse and (scalar) shift perturbations, respectively, and $-\psi\delta_{ij} + \partial_i\partial_jE$ is the scalar part of the linearized spatial metric.

With $\phi = \phi\left(t\right)+\delta\phi\left(t, \textbf{x}\right)$, $\chi = \chi\left(t\right)+\delta\chi\left(t, \textbf{x}\right)$ denoting small inhomogeneities in the scalar fields around the homogeneous background, the linearized action~\eqref{eq:setup_action} takes the form:
\begin{equation}
\begin{aligned}
\label{app:action}
\mathcal{L}
 = a^3 &\Bigg( -3\dot{\psi}^2 + {\textstyle \frac{k^2}{a^2}}\psi^2 
- 2 {\textstyle\frac{k^2}{a^2}}\sigma_{sh}\Big( \dot{\psi} + H\alpha \Big)
-2 \left( 3H\dot{\psi} + {\textstyle \frac{k^2}{a^2}}\psi\right) \alpha
\\
& + \Big(\dot{\phi}\delta\phi + \big(\Sigma_1+\Sigma_2\dot{\chi}^2\big)\, \dot{\chi}\delta\chi\Big)\Big(3\dot{\psi} + {\textstyle \frac{k^2}{a^2}}\sigma_{sh}\Big)
\\
& - \Big(3H^2- {\textstyle \frac12} \dot{\phi}^2 
- {\textstyle \frac12} \big(\Sigma_1+3\Sigma_2\dot{\chi}^2\big) \, \dot{\chi}^2\Big)\,\alpha^2
\\
& - \Big( \dot{\phi}\delta\dot{\phi} 
+ \big( {\textstyle \frac12}\Sigma_{1,\phi}\dot{\chi}^2
+ {\textstyle \frac34}\Sigma_{2,\phi}\dot{\chi}^4 + V_{,\phi} \big)\, \delta\phi 
\Big)\, \alpha
\\
& - \Big( \big(\Sigma_1 + 3\Sigma_2\dot{\chi}^2 \big)\, \dot{\chi}\delta\dot{\chi} 
 + V_{,\chi}\delta\chi\Big)\, \alpha
\\
&+{\textstyle \frac12}\delta\dot{\phi}^2 
- {\textstyle \frac12} {\textstyle \frac{k^2}{a^2}}\delta\phi^2
+{\textstyle \frac12} \Big({\textstyle \frac12} \Sigma_{1,\phi\phi}\dot{\chi}^2
+{\textstyle \frac14}\Sigma_{2,\phi\phi}\dot{\chi}^4 - V_{,\phi\phi} \Big)\delta\phi^2
\\
&+{\textstyle \frac12}\Big(\Sigma_1+3\Sigma_2\dot{\chi}^2\Big)\delta\dot{\chi}^2
-{\textstyle \frac12}\Big(\Sigma_1
+ \Sigma_2\dot{\chi}^2\Big) {\textstyle \frac{k^2}{a^2}} \delta\chi^2
- {\textstyle \frac12}V_{,\chi\chi}\delta\chi^2
\\
& +\left(\Sigma_{1,\phi}\dot{\chi} + \Sigma_{2,\phi}\dot{\chi}^3\right)\delta\phi\delta\dot{\chi}
- V_{,\phi\chi}\delta\phi\delta\chi 
\Bigg)\,,
\end{aligned}
\end{equation}
where  $\sigma_{sh} \equiv a\left(a\dot{E} - \beta\right)$ is the scalar part of the linearized shear. 

Variation of Eq.~\eqref{app:action} with respect to $\alpha$ and $\beta$ leads to the linearized Hamiltonian and momentum constraints:
\begin{subequations}
\label{eq:app_constraints}
\begin{align}
- 2\, {\textstyle \frac{k^2}{a^2}}\Big(\psi + H\sigma_{sh}\Big)
=&  \Big( 6H^2 -\dot{\phi}^2 - \Sigma_1\dot{\chi}^2 - 3\Sigma_2\dot{\chi}^4\Big)\, \alpha 
\\
+& 6H\dot{\psi} +\dot{\phi}\delta\dot{\phi} 
+ \Big(\Sigma_1 + 3\Sigma_2\dot{\chi}^2 \Big)\,\dot{\chi}\delta\dot{\chi}
\nonumber\\
+&\Big( {\textstyle \frac12} \Sigma_{1,\phi}\dot{\chi}^2 
+ {\textstyle \frac34}\Sigma_{2,\phi}\dot{\chi}^4 + V_{,\phi} \Big)\,\delta\phi  +  V_{,\chi}\delta\chi
\,,\nonumber
\\
H\alpha + \dot{\psi} =& {\textstyle \frac12}\left(\dot{\phi}\delta\phi+\left(\Sigma_1
+\Sigma_2\dot{\chi}^2\right)\dot{\chi}\delta\chi\right).
\end{align}
\end{subequations}

Varying Eq.~\eqref{app:action} with respect to $\delta\phi$ and $\delta\chi$ yields the evolution equations for the perturbed scalar fields:
\begin{subequations}
\label{eq:app_eom}
\begin{align}
&\delta\ddot{\phi} + 3H\delta\dot{\phi} 
+ \left( {\textstyle \frac{k^2}{a^2}} + V_{,\phi\phi} - {\textstyle \frac12} \Sigma_{1,\phi\phi} \dot{\chi}^2
- {\textstyle \frac14} \Sigma_{2,\phi\phi}\dot{\chi}^4\right)\delta\phi
\\
- &  \Big( \ddot{\phi} + 3H\dot{\phi} - {\textstyle \frac12} \Sigma_{1,\phi}\dot{\chi}^2 - {\textstyle \frac34}\Sigma_{2,\phi}\dot{\chi}^4 - V_{,\phi} \Big)\,\alpha 
\nonumber\\
- &  \dot{\phi}\, \Big(  \dot{\alpha} + 3 \dot{\psi} + {\textstyle \frac{k^2}{a^2}} \sigma_{sh}  \Big)
- \left(\Sigma_{1,\phi}+\Sigma_{2,\phi}\dot{\chi}^2\right)\dot{\chi}\delta\dot{\chi}
+ V_{,\phi\chi}\delta\chi
=0\,;
\nonumber\\
& \Big(\Sigma_1 + 3\Sigma_2\dot{\chi}^2\Big)\Big( \delta\ddot{\chi}+ 3H\delta\dot{\chi}\Big)
+ \Big(\Sigma_1+3\Sigma_2\dot{\chi}^2\Big)^\textbf{.} \,\delta\dot{\chi}
\\
+& \left(\Sigma_1+\Sigma_2\dot{\chi}^2\right) {\textstyle \frac{k^2}{a^2}} \delta\chi 
+V_{,\chi\chi}\delta\chi
\nonumber\\
-& \Big(\Sigma_1  + 3\Sigma_2\dot{\chi}^2 \Big)\Big(\ddot{\chi}  + 3H\dot{\chi} \Big)\,\alpha
- \Big(\Sigma_1  + 3\Sigma_2\dot{\chi}^2\Big)^\textbf{.}\,\dot{\chi} \alpha
+ V_{,\chi} \, \alpha 
 \nonumber\\
-& \left(\Sigma_1\dot{\chi}  + 3\Sigma_2\dot{\chi}^3\right) \dot{\alpha}
- \left(\Sigma_1 \dot{\chi} + \Sigma_2\dot{\chi}^3\right) \left(3\dot{\psi} + {\textstyle \frac{k^2}{a^2}}\sigma_{sh}\right)
\nonumber\\
+& \left(\Sigma_{1,\phi}+\Sigma_{2,\phi}\dot{\chi}^2\right)\dot{\chi}\delta\dot{\phi}  +V_{,\phi\chi}\delta\phi
\nonumber\\
+& \Big(\Sigma_{1,\phi}+\Sigma_{2,\phi}\dot{\chi}^2\Big) \Big(\ddot{\chi} + 3H \Big) \,\delta\phi 
+ \left(\Sigma_{1,\phi}+\Sigma_{2,\phi}\dot{\chi}^2\right)^\textbf{.}\,\dot{\chi}\delta\phi
=0.
\nonumber
\end{align}
\end{subequations}

In spatially-flat gauge ($\psi, E \equiv 0$), the evolution and constraint equations~(\ref{eq:app_constraints}-\ref{eq:app_eom}) together with the background equations~(\ref{eq:setup_eom_field},\,\ref{eq:setup_eom_gravity}) yield a closed system for the dynamical variables $\delta\phi$ and $\delta\chi$ which we used above to numerically compute the gauge-invariant quantities ${\cal S}$ and ${\cal R}$ defined in Eqs.~\eqref{def-S-hydro} and \eqref{eq:sourcing_R_F}, respectively.

\bibliographystyle{apsrev}
\bibliography{bib_paper2}

\end{document}